# Handheld Haptic Device with Coupled Bidirectional Input

Megh Vipul Doshi[1], Michael Hagenow[1], Robert Radwin[2], Michael Gleicher[3], Bilge Mutlu[3], and Michael Zinn[1]

*Abstract*—Handheld kinesthetic haptic interfaces can provide greater mobility and richer tactile information as compared to traditional grounded devices. In this paper, we introduce a new handheld haptic interface which takes input using bidirectional coupled finger flexion. We present the device design motivation and design details and experimentally evaluate its performance in terms of transparency and rendering bandwidth using a handheld prototype device. In addition, we assess the device's functional performance through a user study comparing the proposed device to a commonly used grounded input device in a set of targeting and tracking tasks.

## I. INTRODUCTION

Kinesthetic haptic devices offer a variety of ways to interact with users, from rendering virtual environments to providing guidance and feedback during teleoperation of robots. Traditionally, high-performance haptic devices have been grounded (e.g., the devices from Force Dimension [1] or Haption [2]), meaning that they are fixed in a location and generate haptic sensations by reacting against the environment. More recently, a variety of handheld haptic devices have been proposed that provide similar kinesthetic renderings by reacting against the user's hand or arm [3]. For example, Dills et al. [4] propose a high-performance one degree-of-freedom device employing hybrid actuation. Many of the other recent handheld devices provide haptic feedback to each finger individually either through finger-mounted devices [5], [6] or gloves [7], [8]. Notably, many of the recent handheld devices are designed for rendering virtual environments in gaming or virtual-reality applications. Alternatively, we are interested in one degree-of-freedom industrial applications and propose a new handheld haptic input device that is actuated using one hand through two mechanically-coupled triggers.

This work was supported in part by a NASA University Leadership Initiative (ULI) grant awarded to the UW-Madison and The Boeing Company (Cooperative Agreement # 80NSSC19M0124) and by DGE - 2152163 (NRT) Integrating Robots into the Future of Work

[1]Megh Vipul Doshi, Michael Hagenow and Michael Zinn are with the Department of Mechanical Engineering, University of Wisconsin–Madison, Madison 53706, USA [megh.doshi|mhagenow|mzinn]@wisc.edu

[2]Robert Radwin is with the Department of Industrial and Systems Engineering, University of Wisconsin–Madison, Madison 53706, USA rradwin@wisc.edu

[3]Michael Gleicher, and Bilge Mutlu are with the Department of Computer Sciences, University of Wisconsin–Madison, Madison 53706, USA [gleicher|bilge]@cs.wisc.edu

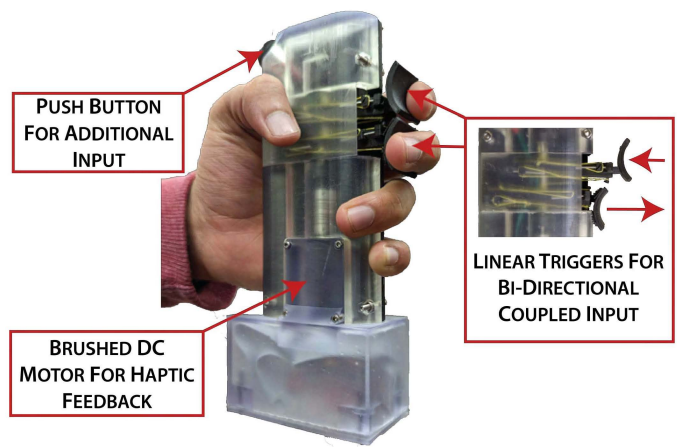

Fig. 1. Our proposed handheld haptic device with two triggers that take input via finger flexion. The triggers are mechanically coupled (i.e., when one trigger is pushed in, the other pushes out).

There are a variety of industrial applications that could benefit from a one degree-of-freedom haptic input device. For example, an operator could precisely control a single variable during an industrial process (e.g., flow rates, temperatures, feed rates, pressure during sanding) and the device could provide haptic cues or guidance as necessary (e.g., vibration, modulation of stiffness). In many applications, it would also be desirable for the input to be differential, meaning an operator may want to adjust a particular process variable from the current set point (e.g., go *faster* or *slower*). Finally, many industrial applications would benefit from a device that is one-handed, which frees the operators' other hand for secondary tasks (e.g., controlling a second input or feeling a surface during sanding). To address these desired qualities, we propose a one degree-of-freedom bidirectional input where the differential input is provided through two-mechanically coupled triggers, as shown in Figure 1. We were inspired by existing applications leveraging finger flexion, such as musical instruments (e.g., a trumpet), that allow for input in the natural direction of the finger. Specifically, we chose to investigate a mechanism leveraging adjacent finger flexion to utilize the innate structure of the human fingers, that is more adept at pulling than pushing [9]. Additionally, the mechanically-coupled triggers allows for simple bidirectional inputs where the triggers correspond to opposite directions.

In this paper, we propose a high-performance one degree-of-freedom (DOF) haptic device that is mobile and can be actuated using a single hand through two fingers. We first discuss our design requirements for realizing a usable and high-performance haptic device followed by describing our developed prototype. Through an experimental evaluation, we demonstrate that the prototype has many desirable characteristics for generating high-quality haptics, such as a high transparency and a high bandwidth. Finally, we investigate the performance of our prototype in terms of giving precise input via a user study comparing the device to a haptic knob, a form factor that is already widely used in industrial settings.

## II. Device Design and Implementation

### A. Design Requirements

Our goal was to design a handheld and mobile alternative to a 1-DOF grounded interface. We propose a solution which is to be actuated using adjacent finger flexion (index finger and middle finger), where the inputs are mechanically coupled as can be seen in Figure 1. The range of motion of the device was designed to span half that of a typical adult male finger flexion motion and be capable of rendering up to 15 N peak force [10] [11]. The device should be transparent(i.e. low friction) and capable of rendering stiffness levels comparable to other high performance kinesthetic haptic interfaces. The device requirements are summarized in the list below.

- Allow for 1-DOF haptic interactions via finger flexion
- Mobility with one handed operation
- High stiffness and high transparency
- A maximum force of 15 N at each trigger
- A stroke length of 15 mm for each trigger

### B. Design Implementation

To achieve high transparency, we incorporated as low an inertia as we could with a simple lightweight design (device weight was 383.40 grams). The device has a 3d printed central handle, with two linear triggers that are used to take input from as well as to give haptic feedback to the user. Our device design has a two finger pushing mechanism and is similar in form factor to the trumpet, a musical instrument which also uses finger flexion motion to take input, but different in that our device uses coupled finger flexion. The central handle is also a means to ground against the kinesthetic feedback from the device. The two triggers on the device are mechanically coupled (i.e., when one trigger is pushed in, the other trigger is automatically pushed out and vice versa) via the cable drive train. The linear triggers have low friction guide rails which hold them in the same axial position and a cable drive train which transmits power from the central shaft to the triggers. The cable drive uses spectra cable which allows for smaller bend radii enabling us to build a more compact design. The cable drive train has terminations on the the triggers as well as on the shaft. The terminations on the triggers can be used to re-tension the drive train in case it is affected by creep in the cable or wear and tear of the plastic parts. The central shaft is connected to an ironless core brushed DC motor via a flexible shaft coupling which corrects for any misalignment between the drivetrain shaft and the motor shaft.

We use a Maxon ironless core brushed DC motor (Model No. 339156) which can provide a maximum continous torque of upto 32.3 mNm. This along with the drive train's reduction gives us a force of up to 15 N at the trigger. The motor also has a 4096 CPT encoder which is used to measure the user's input. The motor was also used to generate haptic sensations at the device triggers. We use Copley Junus JSP 90-20 amplifiers along with a TI C2000 series Piccolo 29069M Launchpad to control the system. The system runs at a 1000 Hz in an application written using simulink in an impedance control mode. We also have a button on the device which can be used to take in an additional input from the user if needed.

## III. System Performance Evaluation

To evaluate the performance of our device, we conducted a series of experiments to evaluate (1) the rendering force and stiffness, (2) the friction (i.e. transparency), and (3) the rendering bandwidth of our prototype device. The experiments and their results are described in the following sections.

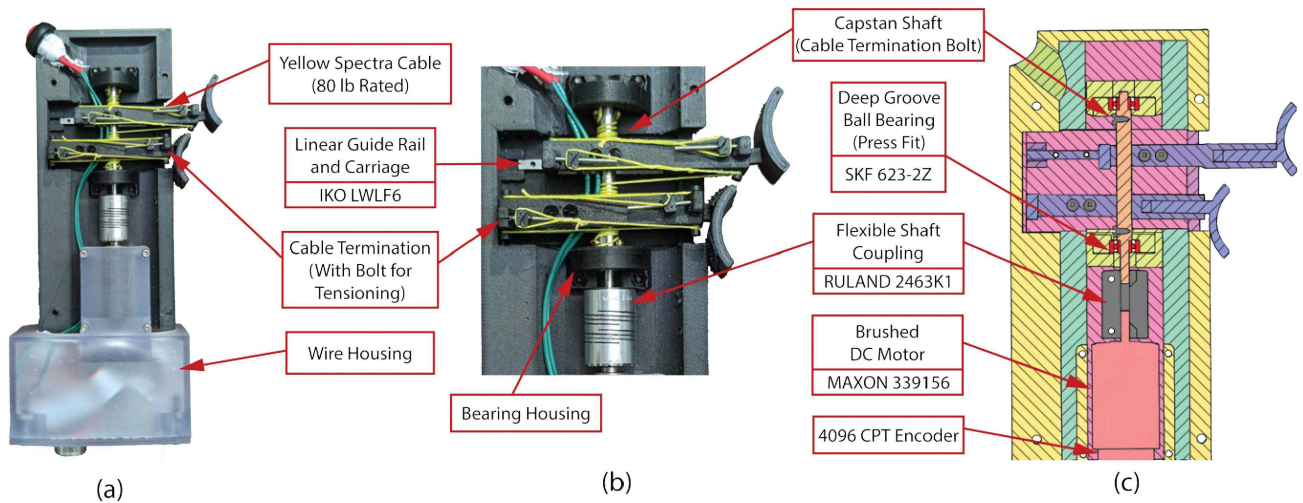

Fig. 2. Internal (a) view, zoomed internal view (b), and cross-sectional view (c) of the device showcasing the drive-train and related design elements.

*A. Maximum Rendering Force and Stiffness*

To assess the device performance, we experimentally evaluated the maximum uncoupled stable rendering stiffness [12], [13]. We perform an experiment wherein we incrementally increase the stiffness rendered by the device in discrete steps and check for stability at each step. Stability is assessed by physically perturbing the finger triggers of the device through user touch and observing whether the system stabilizes. We can qualitatively asses stability by looking at the vibrations in the system i.e. if the vibrations are increasing in amplitude, the system is deemed unstable. To reduce the bias of any single user interaction, we performed the testing with several cohorts. The average maximum stable rendering commanded stiffness was 1.06 N/mm while the maximum stable rendering commanded force was 15.9 N.

*B. Friction tests to assess transparency*

To evaluate the transparency [14], we performed an experiment to measure the friction of the drive train. Specifically, we applied a slowly increasing motor torque and identified the torque level at which the motor and drive-train initiated movement. This experiment was repeated at various motor positions to account for friction variation as a function of device configuration. The commanded motor torques varied from 0.177 to 1.029 mNm, with an average torque of 0.665 mNm. The equivalent (reflected) friction force at the finger interface, taking into account the drive-train reduction, varied from 0.05 to 0.343 N with an average frictional loss of 0.22 N. For comparison, other handheld haptic devices like the CLAW [3] and [15] report frictional losses (or a minimum transparency) of 0.5 N and 0.54 N respectively.

*C. Rendering Bandwidth*

The frequency range over which a haptic device can accurately display forces, with minimal magnitude and phase distortion, is referred to as the rendering bandwidth. A large rendering bandwidth is important for realistic user perception [16]. To achieve a large rendering bandwidth, it is important to design a stiff drive-train such that there are no structural vibration modes present within the desired bandwidth, typically up to 100 Hz for a high performance interface.

To evaluate the rendering bandwidth, we measured the frequency response of the device drive-train. The frequency response was obtained by measuring the motor position output using the 4096 CPT encoder in response to an applied torque chirp signal. A virtual stiffness was overlaid on the torque chirp to maintain centering of the device. To provide a broad-spectrum evaluation, we performed an experiment to measure the frequency response across both low frequency and high frequencies. In the experiment, the disturbance chirp signal was varied from 0.1 - 100 Hz over a 30 second interval. The test was repeated 10 times. The results are shown in Figure 3.

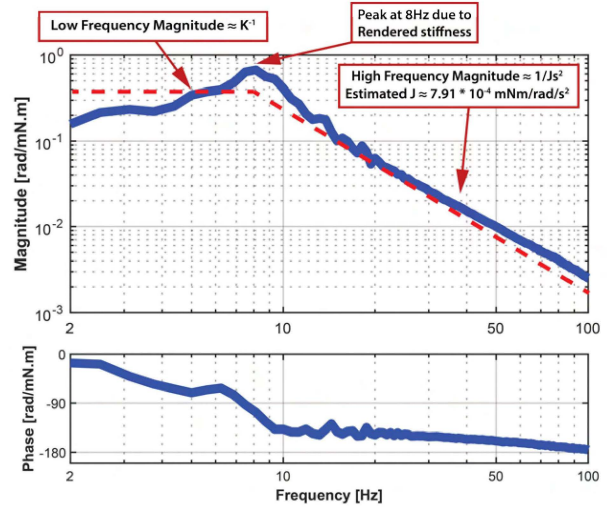

Fig. 3. Frequency response of the proposed device. The only peak in the magnitude plot occurs due to the rendered stiffness.

As seen in Figure 3, the low frequency magnitude response is approximately flat up to 4 Hz, after which it increases as it approaches the resonance created by the introduction of the virtual centering spring. The flat portion of the curve is, as expected, approximately equal to the inverse of the rendered stiffness (0.5 rad/mNm). The high frequency magnitude response shows a peak at approximately 8 Hz, which corresponds to the induced resonance the results from the introduction of the virtual centering stiffness. Importantly, there is no evidence of drive-train flexible modes below 100 Hz. We can estimate the reflected inertia of the device from the high frequency magnitude of the frequency response which is approximately equal to $7.91 \times 10^{-4}$ mNm/rad/s$^2$. A low value of reflected inertia infers highly transparent device.

## IV. USER STUDY

To assess the efficacy of the proposed device, we conducted a user study. Specifically, our study aimed to assess the accuracy and perceived usability of the proposed device across a range of common one degree-of-freedom tasks.

*A. Study Design*

Our study compared our device (referred to as the *handheld* condition going forward) to a grounded rotary input (i.e., knob, see Figure 4) in a within-subjects design where the order of conditions (i.e., the device used for each task) was counterbalanced. We chose the knob as a baseline based on its prevalence in haptics (e.g., DC motors) and society (e.g., control panels, thermostats). Input devices like knobs are typically used to give more precise inputs as suggested by prior studies [17] and standards [18]. In this preliminary evaluation, we focused on using the inputs for two differential control applications: reaching target locations and tracking trajectories. Accordingly, both input devices were programmed to render a static haptic stiffness around a center point, similar to a joystick.

## B. Participants

Our study involved 11 participants (6 male, 5 female), aged 18–23 ($M = 19.27$, $SD = 1.60$) recruited from the university campus. All participants were right handed. None of the participants identified as having extensive experience with haptics. Participants were paid $15 an hour.

## C. Procedure

After providing informed consent, participants were briefed on the structure of the experiment. The participants then completed a targeting task for each condition followed by a tracking task for each condition. The details of the tasks are presented in the next section. After completing each condition, participants filled out the NASA TLX [19] and the System Usability Scale (SUS) questionnaire [20]. The order of the conditions was counterbalanced across participants. For each task, the participants trained with the input device before collecting test data. Following the two tasks, participants completed a brief demographics survey and completed questions on qualitative data about the devices answering questions about their comfort levels with each device for both tasks. The study procedure lasted approximately one hour. The study was administered under a protocol approved by the university Institutional Review Board (IRB).

## D. Apparatus

The knob condition used a Maxon motor (model no. 320165) with a rotary handle attached to the shaft. An encoder attached to the motor was used to measure the user's input. Both conditions were operated as differential inputs using a haptic overlay. For both the tasks, both input devices were programmed to render a static stiffness of 7.5 mNm/rad for the knob and 7.4 mNm/mm for the handheld haptic device which centered the knob and the triggers of the device respectively. A stiffness value that was deemed comfortable for the task was chosen through testing by the authors. This stiffness is what provides the force feedback to the user through the motors in each device, the farther the user moves from the center point of each device, the greater the force. The knob was mapped to move the object on the screen left and right by rotating the knob in those directions. Whereas, the handheld device was mapped to move the object left and right by pushing each trigger in(the upper trigger to move right and lower to move left).

## E. Tasks & Measurements

*1) Targeting Task:* We assessed the targeting performance of each condition through a Fitt's law experiment. In the targeting task, shown in Figure 5, participants used the input device to move a cursor to a designated goal location. The position and size of the goal varied between tests. Participants were instructed to reach the target as fast as possible. When possible, the design of the Fitt's law experiment followed the ISO standard proposed by Mckenzie [21]. In each trial, participants pressed a key to begin and the trial was stopped once the target was reached (defined as when the cursor was within the target for two seconds). For each condition, participants completed 30 training trials and 60 test trials. From each trial, we calculated the index of difficulty ($ID$) and used it to calculate the throughput ($TP$) of the input device.

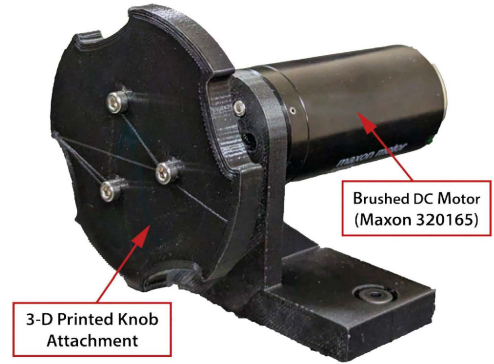

Fig. 4. Knob used as the comparison input device in our study.

$$ID = log_2(\frac{A}{W} + 1)$$
$$TP = \frac{ID}{MT} \quad (1)$$

where $A$ is the amplitude of movement to the target, $W$ is the width of the target, and $MT$ is the time to reach the target. A higher throughput corresponds to better device performance for targeting tasks.

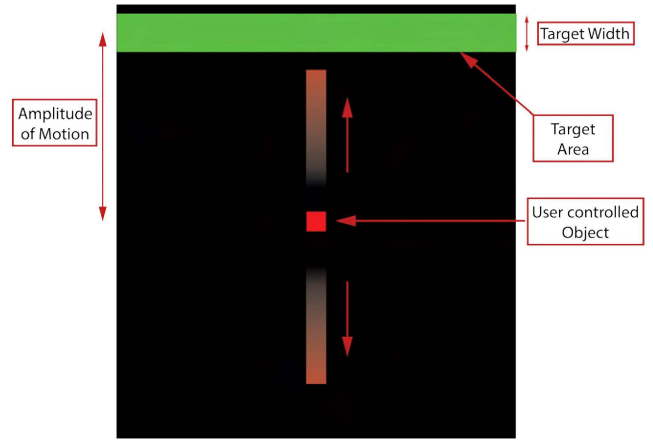

Fig. 5. Targeting task following ISO 9421-9 to calculate throughput of device.

*2) Tracking Task:* We assessed the tracking performance for each condition through a series of sinusoid tracking tests. As shown in Figure 6, participants used the input device to follow the desired trajectory and we tracked the absolute value of error between the reference trajectory and participant cursor at each time-step. The frequency and amplitude of the trajectory were varied during the trials. Participants were instructed to follow the trajectory as closely as possible. To enable direct comparisons of performance across the conditions, we selected two different amplitudes and two different frequencies for the sinusoidal trajectory. The two frequencies, 0.3 Hz and 1 Hz (referred to as the *low frequency* and *high frequency*), were selected within the bandwidth of human control [22]. In the tracking task, each participant performed 16 training trials (i.e., 4 trials of each combination of amplitude and frequency) before completing four test trials. The four trials were completed without any break (e.g., the second sinusoid started immediately upon the completion of the first sinusoid). The order of the effects were randomly generated and different between the training and testing trials. Each sinusoid consisted of 4 full periods of oscillation for the low frequency trajectories and 8 full periods of oscillations for the high frequency trajectories.

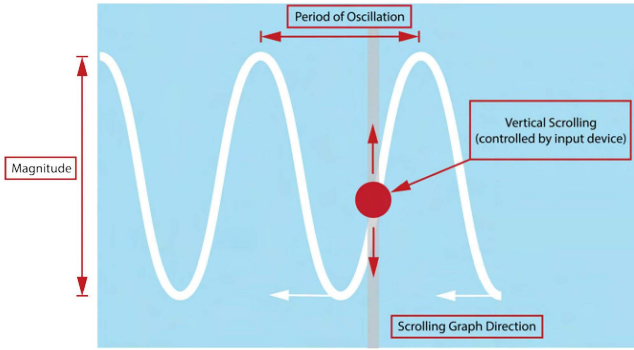

Fig. 6. Task to assess device performance for tracking.

### F. Results

The results for the targeting task were analyzed using a paired two-tailed t-test. We found a significant effect on throughput for the handheld device ($M = 1.65$, $SD = 0.70$) compared to the knob ($M = 2.05$, $SD = 0.60$, $t(10) = 11.48$, $p < 0.05$, $d = 0.59$) with the knob having a higher throughput. The raw NASA TLX score for this task was 30.45 on average for the knob and 38.48 on average for the handheld haptic device. We perform a paired two-tailed T-test on the TLX scores and find that the haptic device ($M = 38.48$, $SD = 16.36$) has significantly higher scores than those for the knob ($M = 30.45$, $SD = 16.36$, $t(10) = 2.28$, $p < 0.05$), .

The error results of the tracking task were analyzed using a three-way repeated measures ANOVA with device (haptic device, knob), frequency (high, low), and amplitude (high, low) as factors. The results are shown in Table I. We find no significant effect of the device on the user error for this task. We do, however, find a significant effect of the amplitude ($F(1,8) = 83.85$, $p < 0.001$) and the frequency ($F(1,8) = 15.57$, $p < 0.001$) of the path on the user error. The average raw NASA TLX score for this task was 42.12 ($M = 42.12$, $SD = 17.42$.) for the knob and 42.42 ($M = 42.42$, $SD = 16.56$) for the handheld haptic device. We perform a paired two-tailed t-test on the TLX scores and find no significant difference ($t(10) = 0.55$, $p = 0.58$) between the scores of this task for both devices. Since the results of the ANOVA do not show any significant relation to the device used, we do not perform any post hoc analysis on this data.

TABLE I
THREE-WAY ANOVA RESULTS

| Independent Variables | F Value | Num of DF | Den DF | PR>F |
|---|---|---|---|---|
| Device | 0.798 | 1.0 | 8.0 | 0.398 |
| Frequency | 19.671 | 1.0 | 8.0 | **<0.05** |
| Amplitude | 62.447 | 1.0 | 8.0 | **<0.001** |
| Device x Frequency | 1.182 | 1.0 | 8.0 | 0.309 |
| Device x Amplitude | 0.011 | 1.0 | 8.0 | 0.920 |
| Frequency x Amplitude | 6.056 | 1.0 | 8.0 | **0.039** |
| Device x Frequency x Amplitude | 0.001 | 1.0 | 8.0 | 0.976 |

TABLE II
AVERAGE ERROR PER SEGMENT

| Trajectory Segment | Device | Mean Error | Standard Deviation |
|---|---|---|---|
| Low Frequency Low Amplitude | Handheld Haptic Device | 33.12 | 15.52 |
| | Knob | 29.19 | 10.76 |
| Low Frequency High Amplitude | Handheld Haptic Device | 81.10 | 25.26 |
| | Knob | 78.18 | 31.86 |
| High Frequency Low Amplitude | Handheld Haptic Device | 51.04 | 13.04 |
| | Knob | 38.17 | 14.85 |
| High Frequency High Amplitude | Handheld Haptic Device | 124.93 | 31.75 |
| | Knob | 113.57 | 55.67 |

The raw average error values per segment are as shown in the Table II below.

The SUS score for our proposed handheld haptic device was 86.60 ($M = 86.60$, $SD = 3.88$) while that of the knob was 87.24 ($M = 87.24$, $SD = 2.57$). This falls in the excellent category in the SUS scale for both devices. We perform a two-tailed paired T-test on the SUS scores as well and find no statistically significant difference between the scores of the two devices ($t(10) : 0.90$, $p = 0.37$).

## V. DISCUSSION

There are several key takeaways from the user study. In the targeting task, we observed that our handheld haptic device did not perform as well as the knob in terms of throughput. Some participants appeared to struggle with the spatial mapping with the handheld haptic device (i.e., which trigger moves the cursor up or down on the screen) and would move the object in the wrong direction before correcting themselves. This effect was not observed when participants used the knob which was both more familiar to participants and grounded to the table in an orientation that was spatially consistent with the task. In the future, we are interested to explore whether the performance with the proposed device could be improved through additional training. The amount of training trials in our experiment was limited to prevent participant fatigue.

In the tracking task, the knob generally had lower average errors than our handheld haptic device for all frequencies and amplitudes. However, the average error varied by less than 13 pixels between the two devices. As seen in Figure 7, much of the difference in performance can be attributed to overshoot by the handheld haptic device, which was not observed when participants used the knob. Looking closely at the high-frequency high amplitude segment of the data in Figure 7, we can see that users commonly undershoot the desired motion when using the knob as compared to our haptic device. However, we see higher overshoots with the haptic device for high-frequency high amplitude trajectories as well. We can infer from these observations that the handheld haptic device may be more suited to tasks where reaching the peaks of the trajectory is more important than controlling the overshoot. The overshoot might also suggest that the haptic overlay was not optimal for the handheld haptic device. In the future, we are interested in exploring the impact of the overlay on user performance.

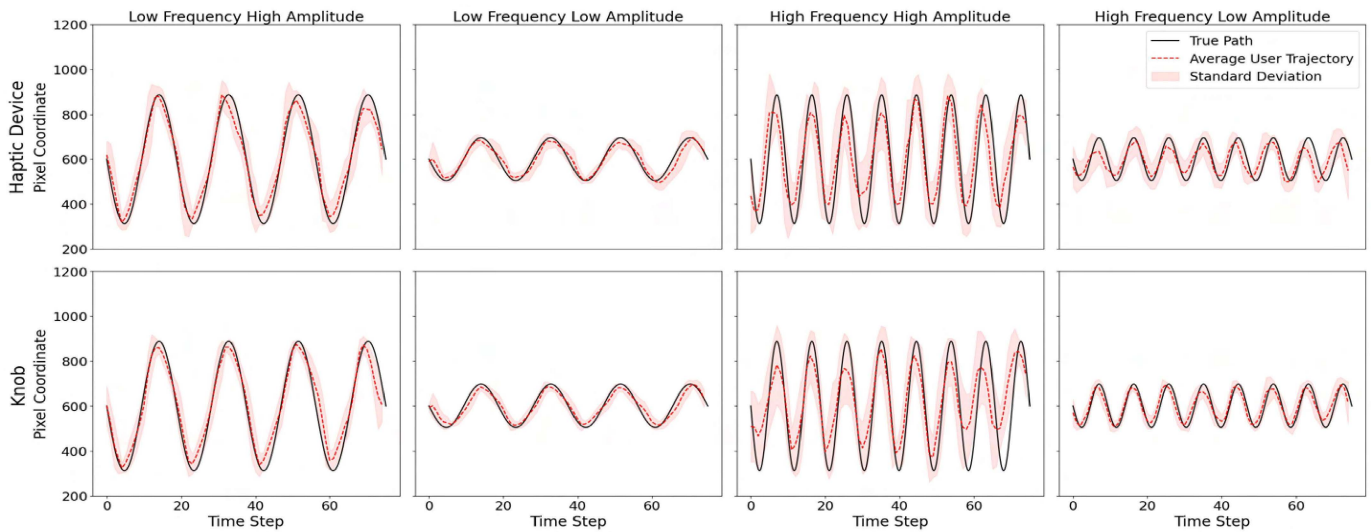

Fig. 7. Average errors and standard deviations for trajectory tracking with the handheld haptic device and the knob.

In terms of usability, both the proposed device and the knob scored in the excellent category for the SUS [20] and the difference between the average usability scores for the devices was only 0.64 percent. Users who play video games frequently (at least once per week) expressed a high degree of satisfcation with the form factor of the handheld haptic device while users who had little or no experience with gaming found the device to be fatiguing.

## VI. Conclusion

In this paper, we designed and evaluated a handheld haptic device that incorporates a bidirectional coupled finger flexion interface. Through experimental validation, we demonstrated that the proposed device can render high-performance haptic effects, as measured by its transparency and rendering bandwidth. Through a user study, we showed that participants rated the proposed device as highly usable and assessed the device with a series of targeting and tracking tasks. While the performance of the device was inferior to that of a haptic knob in some instances, the overall assessment was positive. The results suggest a range of modifications to be considered in future work including improvements to the ergonomic form factor, exploration of alternative haptic effects, and evaluation of these modifications on the overall device and user performance. Specifically, we are interested to test the device in simulated and realistic scenarios. Examples include industrial process control [23] and shared autonomy for collaborative robots [24].

## VII. Acknowledgement

We would like to thank Patrick Dills for his feedback on the mechanical design, Pragathi Praveena for her feedback on study design and Yash Wani for his assistance with programming.